\documentclass[aps,prd,onecolumn,groupedaddress,showpacs,nofootinbib,amssymb]{revtex4}
\usepackage{graphicx,bm,color}
\usepackage{amsmath}
\usepackage{amssymb}
\usepackage{amsfonts}
\usepackage{wrapfig}

\newcommand{\be}{\begin{equation}}
\newcommand{\ee}{\end{equation}}
\newcommand{\bea}{\begin{eqnarray}}
\newcommand{\eea}{\end{eqnarray}}
\newcommand{\beaa}{\begin{eqnarray*}}
\newcommand{\eeaa}{\end{eqnarray*}}

\newcommand{\nn}{\nonumber \\}
\newcommand{\e}{\mathrm{e}}

\allowdisplaybreaks[4]

\begin{document}

\title{A Toy Model of Discretized Gravity in Two Dimensions and its Extentions}

\author{Marcello Rotondo$^{1}$\footnote{
E-mail address: marcello@gravity.phys.nagoya-u.ac.jp}, 
Shin'ichi Nojiri$^{1, 2,}$\footnote{E-mail address:
nojiri@phys.nagoya-u.ac.jp}
}

\affiliation{
$^1$ Department of Physics, Nagoya University, Nagoya
464-8602, Japan \\
$^2$ Kobayashi-Maskawa Institute for the Origin of Particles and
the Universe, Nagoya University, Nagoya 464-8602, Japan}

\begin{abstract}

We propose a toy model of quantum gravity in two dimensions with 
Euclidean signature.
The model is given by a kind of discretization which is different from the 
dynamical triangulation.
We show that there exists a continuum limit and we can calculate 
some physical quantities such as the expectation value of the area, that is, 
the volume of the two dimensional Euclidean space-time. 
We also consider the extensions of the model to higher dimensions.

\end{abstract}


\maketitle

\section{Introduction}

The quantization of gravity could be one of the most important problems 
in physics in the twenty-first century. 
There is a long history in the discretized or lattice formulations of quantum 
gravity.
Especially the dynamical triangulation in two dimensional gravity 
\cite{David:1984tx,David:1985nj,David:1988hj} achieved a great success 
and the formulation was elucidated by using the matrix model 
\cite{Douglas:1989ve,Gross:1989vs,Brezin:1990rb}.
Because the two dimensional gravity coupled with matter could be identified with 
non-critical string theories, models with space-time supersymmetry have 
been also proposed in 
\cite{Marinari:1990fa,Nojiri:1990ud,Nojiri:1990jv,Nojiri:1992zu}.

In this letter, we propose a new formulation of the quantum gravity in two 
dimensions by using a discretization which is different from the dynamical 
triangulation.
We start with one plaquette, which is a square.
Next we consider a diagram where we replace the plaquette with four plaquettes.
As a next step, we include the diagrams where one of the above four plaquettes is replaced with four plaquettes. 
By iterating the replacements, we obtain a fractal discretization of the original 
square as in  Fig.~\ref{Fig1}. 
We show that the continuum limit exists in the model, where we can calculate 
some physical quantities as the expectation value of the area.
The extensions of the model to higher dimensions could be possible as well.

\section{Fractal Discretization}

In two dimensions, we can always choose the conformal gauge where the 
metric $g_{ij}$ can be expressed as 
\be
\label{DLQG1}
g_{ij} = \e^\phi \delta_{ij}\, .
\ee
Here $\delta_{ij}$ is the standard Kronecker delta and $\e^\phi$ 
is the conformal factor. 
Then in two dimensions the infinitesimal area $dS$ can be expressed as 
\be
\label{DLQG2}
dS = \sqrt{g} d^2 x = \e^\phi d^2 x\, .
\ee
Here $g$ is the determinant of $g_{ij}$. 
In two dimensions, the Einstein-Hilbert term is the total derivative, which gives 
the Euler number, and therefore there is no dynamical meaning 
when we require a fixed topology for the space-time. 
Then only the volume or area in two dimensions has a dynamical meaning. 
\begin{figure}[h!]
\centering
\unitlength=0.4mm
\begin{picture}(150,150)
\thicklines

\put(11,11){\line(1,0){128}}
\put(11,11){\line(0,1){128}}
\put(11,139){\line(1,0){128}}
\put(139,11){\line(0,1){128}}

\put(11,75){\line(1,0){128}}
\put(75,11){\line(0,1){128}}

\put(11,43){\line(1,0){128}}
\put(43,11){\line(0,1){128}}

\put(11,107){\line(1,0){128}}
\put(107,11){\line(0,1){128}}

\put(11,27){\line(1,0){128}}
\put(27,11){\line(0,1){128}}

\put(11,91){\line(1,0){128}}
\put(91,11){\line(0,1){128}}

\put(11,123){\line(1,0){128}}
\put(123,11){\line(0,1){128}}

\put(11,59){\line(1,0){32}}
\put(59,11){\line(0,1){32}}

\put(75,59){\line(1,0){64}}
\put(59,75){\line(0,1){64}}

\put(27,19){\line(1,0){64}}
\put(35,11){\line(0,1){16}}
\put(51,11){\line(0,1){16}}
\put(67,11){\line(0,1){16}}
\put(83,11){\line(0,1){16}}

\put(91,83){\line(1,0){32}}
\put(91,99){\line(1,0){32}}
\put(91,115){\line(1,0){16}}
\put(91,131){\line(1,0){16}}
\put(99,75){\line(0,1){64}}
\put(115,75){\line(0,1){32}}

\put(27,115){\line(1,0){16}}
\put(35,107){\line(0,1){16}}

\put(107,35){\line(1,0){32}}
\put(115,27){\line(0,1){32}}

\put(107,51){\line(1,0){16}}
\put(131,27){\line(0,1){16}}

\put(115,39){\line(1,0){8}}
\put(119,35){\line(0,1){8}}

\end{picture}

\caption{An example of the inhomogeneous lattice. 
}\label{Fig1}
\end{figure}
We would like to obtain the two dimensional manifold as the limit of 
an inhomogeneous lattice as in Fig.~\ref{Fig1}, where we identify the number of 
the plaquettes with the area of the manifold. 
Then, the conformal factor should be proportional to the density of the 
plaquettes.

We may randomly generate the diagram of the inhomogeneous lattice by 
the iteration where one plaqutte is replaced by four plaquettes as 
in Fig.~\ref{Fig2}. 
\begin{figure}[h!]
\centering

\unitlength=0.5mm
\begin{picture}(330,50)
\thicklines

\put(14,14){\line(1,0){32}}
\put(14,14){\line(0,1){32}}
\put(14,46){\line(1,0){32}}
\put(46,14){\line(0,1){32}}

\put(60,30){\makebox(0,0){$\Rightarrow$}}
\put(30,5){\makebox(0,0){Step 0}}

\put(74,14){\line(1,0){32}}
\put(74,14){\line(0,1){32}}
\put(74,46){\line(1,0){32}}
\put(106,14){\line(0,1){32}}
\put(74,30){\line(1,0){32}}
\put(90,14){\line(0,1){32}}

\put(120,30){\makebox(0,0){$\Rightarrow$}}
\put(90,5){\makebox(0,0){Step 1}}

\put(134,14){\line(1,0){32}}
\put(134,14){\line(0,1){32}}
\put(134,46){\line(1,0){32}}
\put(166,14){\line(0,1){32}}
\put(134,30){\line(1,0){32}}
\put(150,14){\line(0,1){32}}

\put(134,38){\line(1,0){16}}
\put(142,30){\line(0,1){16}}

\put(174,14){\line(1,0){32}}
\put(174,14){\line(0,1){32}}
\put(174,46){\line(1,0){32}}
\put(206,14){\line(0,1){32}}
\put(174,30){\line(1,0){32}}
\put(190,14){\line(0,1){32}}

\put(190,38){\line(1,0){16}}
\put(198,30){\line(0,1){16}}

\put(210,5){\makebox(0,0){Step 2}}

\put(214,14){\line(1,0){32}}
\put(214,14){\line(0,1){32}}
\put(214,46){\line(1,0){32}}
\put(246,14){\line(0,1){32}}
\put(214,30){\line(1,0){32}}
\put(230,14){\line(0,1){32}}

\put(230,22){\line(1,0){16}}
\put(238,14){\line(0,1){16}}

\put(254,14){\line(1,0){32}}
\put(254,14){\line(0,1){32}}
\put(254,46){\line(1,0){32}}
\put(286,14){\line(0,1){32}}
\put(254,30){\line(1,0){32}}
\put(270,14){\line(0,1){32}}

\put(254,22){\line(1,0){16}}
\put(262,14){\line(0,1){16}}

\end{picture}

\unitlength=0.5mm
\begin{picture}(130,50)
\thicklines

\put(0,30){\makebox(0,0){$\Rightarrow$}}

\put(70,5){\makebox(0,0){Step 3}}

\put(14,14){\line(1,0){32}}
\put(14,14){\line(0,1){32}}
\put(14,46){\line(1,0){32}}
\put(46,14){\line(0,1){32}}
\put(14,30){\line(1,0){32}}
\put(30,14){\line(0,1){32}}

\put(30,22){\line(1,0){16}}
\put(38,14){\line(0,1){16}}

\put(14,38){\line(1,0){16}}
\put(22,30){\line(0,1){16}}

\put(54,14){\line(1,0){32}}
\put(54,14){\line(0,1){32}}
\put(54,46){\line(1,0){32}}
\put(86,14){\line(0,1){32}}
\put(54,30){\line(1,0){32}}
\put(70,14){\line(0,1){32}}

\put(54,22){\line(1,0){16}}
\put(62,14){\line(0,1){16}}

\put(62,26){\line(1,0){8}}
\put(66,22){\line(0,1){8}}

\put(110,30){\makebox(0,0){$+\ \cdots$}}

\end{picture}

\caption{An example of the inhomogeneous lattice. 
}\label{Fig2}
\end{figure}
Summing up the contributions from all the lattice diagrams may correspond 
to the quantum corrections. 
We may consider the partition function $Z$ defined by 
\be
\label{DLQG3}
Z(x) \equiv \sum_{n=1}^\infty A_n x^n \, .
\ee
Here $n$ is the number of the plaquettes and $A_n$ is the number of 
the diagrams made of $n$ plaquettes generated by the iterative procedure above.
If we write $x$ as $x=\e^{-\lambda}$ with a parameter $\lambda$, 
we obtain $x^n = \e^{-\lambda n}$. 
Because we identify $n$ with the area of the space-time, the parameter 
$\lambda$ can be regarded as a bare cosmological constant. 
If we define 
\be
\label{DLQG3b}
z \equiv Z - x\, ,
\ee
we find a recursion relation as in the following picture
\be
\label{DLQGf}
\unitlength=0.5mm
\begin{picture}(260,50)
\thicklines

\put(2,22){\line(1,0){16}}
\put(2,22){\line(0,1){16}}
\put(2,38){\line(1,0){16}}
\put(18,22){\line(0,1){16}}

\put(10,30){\makebox(0,0){$z$}}

\put(25,30){\makebox(0,0){$=$}}

\put(32,22){\line(1,0){16}}
\put(32,22){\line(0,1){16}}
\put(32,38){\line(1,0){16}}
\put(48,22){\line(0,1){16}}
\put(32,30){\line(1,0){16}}
\put(40,22){\line(0,1){16}}

\put(55,30){\makebox(0,0){$+$}}

\put(62,12){\line(1,0){16}}
\put(62,12){\line(0,1){16}}
\put(62,28){\line(1,0){16}}
\put(78,12){\line(0,1){16}}
\put(62,20){\line(1,0){16}}
\put(70,12){\line(0,1){16}}

\put(66,16){\makebox(0,0){$z$}}

\put(62,32){\line(1,0){16}}
\put(62,32){\line(0,1){16}}
\put(62,48){\line(1,0){16}}
\put(78,32){\line(0,1){16}}
\put(62,40){\line(1,0){16}}
\put(70,32){\line(0,1){16}}

\put(66,44){\makebox(0,0){$z$}}

\put(82,12){\line(1,0){16}}
\put(82,12){\line(0,1){16}}
\put(82,28){\line(1,0){16}}
\put(98,12){\line(0,1){16}}
\put(82,20){\line(1,0){16}}
\put(90,12){\line(0,1){16}}

\put(94,16){\makebox(0,0){$z$}}

\put(82,32){\line(1,0){16}}
\put(82,32){\line(0,1){16}}
\put(82,48){\line(1,0){16}}
\put(98,32){\line(0,1){16}}
\put(82,40){\line(1,0){16}}
\put(90,32){\line(0,1){16}}

\put(94,44){\makebox(0,0){$z$}}

\put(105,30){\makebox(0,0){$+$}}

\put(112,12){\line(1,0){16}}
\put(112,12){\line(0,1){16}}
\put(112,28){\line(1,0){16}}
\put(128,12){\line(0,1){16}}
\put(112,20){\line(1,0){16}}
\put(120,12){\line(0,1){16}}

\put(116,16){\makebox(0,0){$z$}}
\put(116,24){\makebox(0,0){$z$}}

\put(112,32){\line(1,0){16}}
\put(112,32){\line(0,1){16}}
\put(112,48){\line(1,0){16}}
\put(128,32){\line(0,1){16}}
\put(112,40){\line(1,0){16}}
\put(120,32){\line(0,1){16}}

\put(116,44){\makebox(0,0){$z$}}
\put(124,44){\makebox(0,0){$z$}}

\put(132,12){\line(1,0){16}}
\put(132,12){\line(0,1){16}}
\put(132,28){\line(1,0){16}}
\put(148,12){\line(0,1){16}}
\put(132,20){\line(1,0){16}}
\put(140,12){\line(0,1){16}}

\put(144,16){\makebox(0,0){$z$}}
\put(136,24){\makebox(0,0){$z$}}

\put(132,32){\line(1,0){16}}
\put(132,32){\line(0,1){16}}
\put(132,48){\line(1,0){16}}
\put(148,32){\line(0,1){16}}
\put(132,40){\line(1,0){16}}
\put(140,32){\line(0,1){16}}

\put(144,44){\makebox(0,0){$z$}}
\put(136,36){\makebox(0,0){$z$}}

\put(152,12){\line(1,0){16}}
\put(152,12){\line(0,1){16}}
\put(152,28){\line(1,0){16}}
\put(168,12){\line(0,1){16}}
\put(152,20){\line(1,0){16}}
\put(160,12){\line(0,1){16}}

\put(156,16){\makebox(0,0){$z$}}
\put(164,16){\makebox(0,0){$z$}}

\put(152,32){\line(1,0){16}}
\put(152,32){\line(0,1){16}}
\put(152,48){\line(1,0){16}}
\put(168,32){\line(0,1){16}}
\put(152,40){\line(1,0){16}}
\put(160,32){\line(0,1){16}}

\put(164,44){\makebox(0,0){$z$}}
\put(164,36){\makebox(0,0){$z$}}

\put(175,30){\makebox(0,0){$+$}}

\put(182,12){\line(1,0){16}}
\put(182,12){\line(0,1){16}}
\put(182,28){\line(1,0){16}}
\put(198,12){\line(0,1){16}}
\put(182,20){\line(1,0){16}}
\put(190,12){\line(0,1){16}}

\put(186,16){\makebox(0,0){$z$}}
\put(186,24){\makebox(0,0){$z$}}
\put(194,16){\makebox(0,0){$z$}}

\put(182,32){\line(1,0){16}}
\put(182,32){\line(0,1){16}}
\put(182,48){\line(1,0){16}}
\put(198,32){\line(0,1){16}}
\put(182,40){\line(1,0){16}}
\put(190,32){\line(0,1){16}}

\put(186,44){\makebox(0,0){$z$}}
\put(186,36){\makebox(0,0){$z$}}
\put(194,44){\makebox(0,0){$z$}}

\put(202,12){\line(1,0){16}}
\put(202,12){\line(0,1){16}}
\put(202,28){\line(1,0){16}}
\put(218,12){\line(0,1){16}}
\put(202,20){\line(1,0){16}}
\put(210,12){\line(0,1){16}}

\put(214,16){\makebox(0,0){$z$}}
\put(214,24){\makebox(0,0){$z$}}
\put(206,16){\makebox(0,0){$z$}}

\put(202,32){\line(1,0){16}}
\put(202,32){\line(0,1){16}}
\put(202,48){\line(1,0){16}}
\put(218,32){\line(0,1){16}}
\put(202,40){\line(1,0){16}}
\put(210,32){\line(0,1){16}}

\put(214,44){\makebox(0,0){$z$}}
\put(206,44){\makebox(0,0){$z$}}
\put(214,36){\makebox(0,0){$z$}}

\put(225,30){\makebox(0,0){$+$}}

\put(232,22){\line(1,0){16}}
\put(232,22){\line(0,1){16}}
\put(232,38){\line(1,0){16}}
\put(248,22){\line(0,1){16}}
\put(232,30){\line(1,0){16}}
\put(240,22){\line(0,1){16}}

\put(236,26){\makebox(0,0){$z$}}
\put(236,34){\makebox(0,0){$z$}}
\put(244,26){\makebox(0,0){$z$}}
\put(244,34){\makebox(0,0){$z$}}

\end{picture}
\, .
\ee
Then we find
\be
\label{DLQG4}
z = x^4 + 4 x^3 z + 6 x^2 z^2 +4 x z^3 + z^4 = \left( x + z \right)^4\, .
\ee
If we solve Eq.~(\ref{DLQG4}) iteratively by assuming $x$ is small, we obtain 
the sum of inhomogeneous lattices as in (\ref{DLQG3}) or Fig.~\ref{Fig2}. 
By using the standard Ferrari formula for the algebraic equations, we can 
explicitly find that the solution of (\ref{DLQG4}) for $z$, which 
has the following form:
\be
\label{DLQG5}
z= - x + \sqrt{\frac{a}{2}} - \sqrt{ - \frac{a}{2} + \frac{1}{2 \sqrt{2 a}}}\, ,
\ee
that is, 
\be
\label{DLQG5B}
Z= \sqrt{\frac{a}{2}} \left( 1 - \sqrt{ \frac{1}{\sqrt{2 a^3}} -1 } \right)\, .
\ee
Here $a$ is given by
\be
\label{DLQG6}
a = \frac{1}{2} \left\{ \left( \frac{1}{2} + \frac{1}{2} 
\sqrt{ 1 - 4\left(\frac{4}{3} x\right)^3} \right)^{\frac{1}{3}}
+ \left( \frac{1}{2} - \frac{1}{2} \sqrt{ 1 - 4\left(\frac{4}{3} x\right)^3} 
\right)^{\frac{1}{3}} \right\} \, .
\ee
Here we have chosen the branch of the solution so that the Taylor expansion 
with respect to $x$ of $z$ reproduces the perturbative series in (\ref{DLQG4}). 
The explicit values of $A_n$'s in (\ref{DLQG3}) are given in Appendix \ref{appendix}. 
The general solution of (\ref{DLQG5}) is given by 
\be
\label{DLQG7}
z= - x + \epsilon_1 \sqrt{\frac{a}{2}} - \epsilon_2 \sqrt{ - \frac{a}{2} 
+ \frac{\epsilon_1 }{2 \sqrt{2 a}}}\, , \quad \epsilon_{1,2} = \pm 1\, .
\ee
When $\epsilon_1=-1$, the solutions become imaginary. 
Eq.~(\ref{DLQG4}) tells that $z$ in (\ref{DLQG5}) is given by the intersection 
points of two curves, $y=z$ and $y=\left( x + z \right)^4$ as in Fig.~\ref{Fig3}. 
Then, as given in the picture below, there is a critical value $x_c$ of $x$, where 
if $x<x_c$ there are two real solutions and if $x>x_c$ there is no real solution. 
When $x<x_c$, the smaller solution corresponds to the solution 
in (\ref{DLQG5}).  
\begin{figure}[h!]
\centering

\unitlength=0.5mm
\begin{picture}(180,60)
\thinlines

\put(0,20){\vector(1,0){45}}
\put(25,10){\vector(0,1){40}}
\put(50,20){\makebox(0,0){$z$}}
\put(25,55){\makebox(0,0){$y$}}

\thicklines
\put(15,10){\line(1,1){30}}
\qbezier(5,25)(2,30)(0,50)
\qbezier(20,20)(10,20)(5,25)
\qbezier(20,20)(30,20)(35,25)
\qbezier(35,25)(38,30)(40,50)

\put(55,40){\makebox(0,0){$y=z$}}
\put(50,55){\makebox(0,0){$y=\left(x + z\right)^4$}}

\put(25,5){\makebox(0,0){$x<x_c$}}

\thinlines

\put(60,20){\vector(1,0){45}}
\put(85,10){\vector(0,1){40}}
\put(110,20){\makebox(0,0){$z$}}
\put(85,55){\makebox(0,0){$y$}}

\thicklines
\put(75,10){\line(1,1){30}}
\qbezier(75,20)(65,20)(60,25)
\qbezier(75,20)(85,20)(90,25)
\qbezier(90,25)(93,30)(95,50)  

\put(85,5){\makebox(0,0){$x=x_c$}}

\thinlines

\put(120,20){\vector(1,0){45}}
\put(145,10){\vector(0,1){40}}
\put(170,20){\makebox(0,0){$z$}}
\put(145,55){\makebox(0,0){$y$}}

\thicklines
\put(135,10){\line(1,1){30}}
\qbezier(120,20)(130,20)(135,25)
\qbezier(135,25)(138,30)(140,50)  

\put(145,5){\makebox(0,0){$x>x_c$}}

\end{picture}
\caption{The intersection 
points of two curves, $y=z$ and $y=\left( x + z \right)^4$. 
The middle figure corresponds to the critical point $x=x_c$. 
}\label{Fig3}
\end{figure}
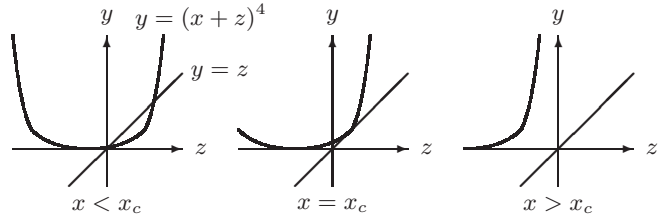
When $x=x_c$, we find $ - \frac{a}{2} + \frac{1}{2 \sqrt{2 a}} =0$ 
in (\ref{DLQG5}) and $1 - 4 (\frac{4}{3} x)^3=0$ in (\ref{DLQG6}), that is, 
\be
\label{DLQG8}
x_c = \frac{3}{4 \cdot 4^{\frac{1}{3}}}\, .
\ee
When $x=x_c$, we find 
\be
\label{DLQG8b}
a = a_c \equiv \frac{1}{2^{\frac{1}{3}}}\, .
\ee
We now consider the meaning of the critical value in (\ref{DLQG8}). 
Because $x=\e^{-\lambda}$, we find that the expectation value 
$\left< A \right>$ of the area $A$ of the manifold is given by 
\begin{align}
\label{DLQG9}
\left< A \right> = & - \frac{d \ln Z}{d\lambda} 
= - \frac{1}{Z} \left\{ \frac{1}{2\sqrt{2a}} + \frac{\frac{1}{4} 
+ \frac{1}{8 \sqrt{ 2 a^3}}}{\sqrt{- \frac{a}{2} + \frac{1}{2 \sqrt{2 a}}}} 
\right\}\frac{da}{d\lambda} \, , \nn
\frac{da}{d\lambda} = & \frac{1}{6} \left\{ \frac{1}{\left( \frac{1}{2} 
+ \frac{1}{2} \sqrt{ 1 -4 \left(\frac{4}{3} x\right)^3} \right)^{\frac{1}{3}}} 
 -   \frac{1}{\left( \frac{1}{2} - \frac{1}{2} \sqrt{ 1 - 4 \left(\frac{4}{3} x\right)^3} 
\right)^{\frac{1}{3}}} \right\} 
\frac{ \frac{64}{9}x^3}{\sqrt{ 1 - 4 \left(\frac{4}{3} x\right)^3}}\, .
\end{align}
Then, in the limit $x\to x_c$, we find
\be
\label{DLQG10}
Z \to \frac{1}{2^{\frac{2}{3}}}\, , \quad \frac{da}{d\lambda} 
\to - \frac{4\cdot 2^{\frac{1}{3}}}{27}\, ,
\ee
and 
\be
\label{DLQG11}
\left< A \right> \to \frac{\gamma}{
\left( \lambda - \lambda_c \right)}\, , \quad \lambda_c \equiv - \ln x_c \, ,
\ee
where for simplicity we have defined a parameter 
$\gamma\equiv 2 \cdot 2^{\frac{1}{3}} / 9 \simeq 0.28$. 
We should note that $\left< A \right> $ diverges at $\lambda = \lambda_c$. 
In the continuum limit, the number of the plaquettes becomes infinity, therefore 
the limit $x\to x_c$ or $\lambda \to \lambda_c$ corresponds to the 
continuum limit. 
We can define a renormalized cosmological constant $\lambda_R$ by 
introducing a new parameter $a_L$ which corresponds to the lattice spacing, 
as follows
\be
\label{DLQG12}
\lambda \equiv \lambda_c + a_L^2 \lambda_R\, ,
\ee
and defined the renormalized area of the manifold by 
\be
\label{DLQG13}
\left< A \right>_R \equiv - \frac{d \ln Z}{d\lambda_R}\, ,
\ee
and consider the limit $a_L\to 0$.
Then we find
\be
\label{DLQG14}
\left< A \right>_R \to \frac{\gamma}{\lambda_R}\, .
\ee
We may also evaluate the fluctuation of the area $\delta A$, 
\be
\label{DLQG14b}
\delta A \equiv \sqrt{ \left< A^2 \right>_R - \left< A \right>_R^2 } 
= \sqrt{ \frac{\partial^2 \ln Z}{\partial \lambda_R} }
= \sqrt{ - \frac{\partial \left< A^2 \right>_R^2}{\partial \lambda_R}}
= \frac{\gamma^{\frac{1}{2}}}{\lambda_R} \, .
\ee
We should note that $\delta A > \left< A \right>_R$, which tells that the two dimensional 
surface fluctuates strongly, which is consistent with the results from 
the dynamical triangulation. 

By regarding $Z$ as a thermodynamical partition function, 
we may identify $\lambda_R$ as the inverse of thermodynamical temperature, 
\be
\label{lambdaTbeta}
\lambda_R = \beta = \frac{1}{T}\, .
\ee
Here we set the Boltzmann constant to be unity. 
Eqs.~(\ref{DLQG13}) and (\ref{DLQG14}) tell that the Gibbs free energy $F$
should be given by 
\be
\label{Gibbs}
F \equiv - T \ln Z = \frac{\gamma}{\lambda_R} 
\ln \frac{\lambda_R}{\mu^2} \, ,
\ee
where $\mu$ is a constant. 
In turn, the entropy $S$ is given by 
\be
\label{entropy}
S \equiv \frac{E - F}{T} 
= \gamma \left( 1 - \ln \frac{\lambda_R}{\mu^2} \right) 
= \gamma \ln \left( 
\frac{ \e \mu^2 }{\gamma} \left< A \right>_R \right) \, .
\ee
Then the entropy is given by the logarithmic function of the area rescaled 
by a constant. 
In two space-time dimensions, the area $\left< A \right>_R$ corresponds 
the space-time volume. 
The volume of a time slice could be given by the length and the 
boundaries of the time slice are given by the points. 
Therefore the area in the four dimensional space-time corresponds 
to the points in the two dimensional space-time, whose volume should 
vanish. 
In the four dimensional space-time, the entropy is proportional to the area. 
Then if the entropy is proportional to the `area' even in two dimensional space-time, 
the entropy should vanish. 
Eq.~(\ref{entropy}) could tell that there is a logarithmic correction to the 
entropy in addition to the part proportional to the area. 

In thermodynamics, we can choose two independent variables from 
the three macroscopic variables, that is, the temperature $T$, the pressure $P$, and 
the volume $V$.
Because we are considering the two dimensional surface, the volume $V$ 
is nothing but the area. 
Eqs.~(\ref{DLQG14}) and (\ref{lambdaTbeta}) tell that the renormalized area 
$\left< A \right>_R$ is proportional to the temperature $T$
\be
\label{AreaT}
\left< A \right>_R = \gamma T\, .
\ee
Then the area, which is the volume in two dimensions, and the temperature 
$T$ are not independent from each other. 
Defining $P$ as in standard thermodynamics, we find
\be
\label{pressure}
P = T \frac{\partial S}{\partial \left< A \right>_R} 
= \frac{\gamma T}{\left< A \right>_R} 
= 1\, , 
\ee
which is a constant. 


\section{Possible Extensions}

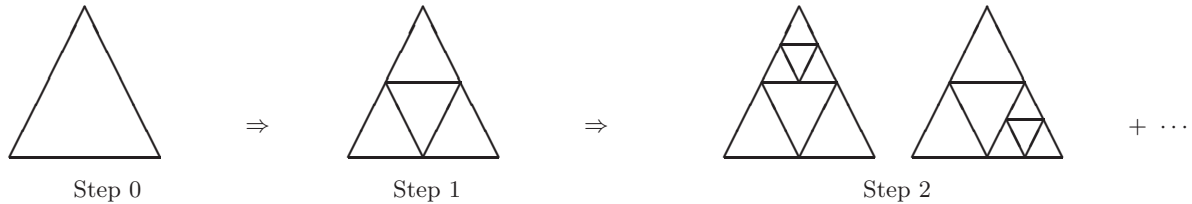
\begin{figure}[h!]
\centering
	
\unitlength=0.5mm
\begin{picture}(330,70)
\thicklines

\put(14,14){\line(1,0){40}}
\put(14,14){\line(1,2){20}}
\put(54,14){\line(-1,2){20}}

\put(80,22){\makebox(0,0){$\Rightarrow$}}
\put(40,5){\makebox(0,0){Step 0}}

\put(104,14){\line(1,0){40}}
\put(104,14){\line(1,2){20}}
\put(144,14){\line(-1,2){20}}

\put(114,34){\line(1,0){20}}
\put(124,14){\line(1,2){10}}
\put(124,14){\line(-1,2){10}}

\put(170,22){\makebox(0,0){$\Rightarrow$}}
\put(125,5){\makebox(0,0){Step 1}}

\put(204,14){\line(1,0){40}}
\put(204,14){\line(1,2){20}}
\put(244,14){\line(-1,2){20}}

\put(214,34){\line(1,0){20}}
\put(224,14){\line(1,2){10}}
\put(224,14){\line(-1,2){10}}	

\put(219,44){\line(1,0){10}}
\put(224,34){\line(1,2){5}}
\put(224,34){\line(-1,2){5}}

\put(254,14){\line(1,0){40}}
\put(254,14){\line(1,2){20}}
\put(294,14){\line(-1,2){20}}

\put(264,34){\line(1,0){20}}
\put(274,14){\line(1,2){10}}
\put(274,14){\line(-1,2){10}}

\put(279,24){\line(1,0){10}}
\put(284,14){\line(1,2){5}}
\put(284,14){\line(-1,2){5}}

\put(250,5){\makebox(0,0){Step 2}}

\put(320,22){\makebox(0,0){$+\ \cdots$}}

\end{picture}

\caption{Series of the triangle lattice.
\label{FigTri}}
\end{figure}

In this letter we considered a specific method of subdividing the initial surface,
but different procedures may be considered. 
Instead of discretizing the surface by using squares, we may consider the recursive 
triangulation shown in Fig.~\ref{FigTri}. 
Because there is a net increment of three plaquettes at each step, 
the recursion formula is identical with Eq.~(\ref{DLQG4}) and the global properties 
are left unchanged.

\begin{figure}[h!]
\centering

\unitlength=0.5mm
\begin{picture}(90,30)
\thicklines

\put(0,0){\line(1,0){20}}
\put(0,0){\line(0,1){20}}
\put(0,20){\line(1,0){20}}
\put(20,0){\line(0,1){20}}
\put(0,20){\line(2,1){10}}
\put(20,0){\line(2,1){10}}
\put(20,20){\line(2,1){10}}
\put(10,25){\line(1,0){20}}
\put(30,5){\line(0,1){20}}

\put(40,10){\makebox(0,0){$\Rightarrow$}}

\put(50,0){\line(1,0){20}}
\put(50,0){\line(0,1){20}}
\put(50,20){\line(1,0){20}}
\put(70,0){\line(0,1){20}}
\put(50,20){\line(2,1){10}}
\put(70,0){\line(2,1){10}}
\put(70,20){\line(2,1){10}}
\put(60,25){\line(1,0){20}}
\put(80,5){\line(0,1){20}}

\put(50,10){\line(1,0){20}}
\put(60,0){\line(0,1){20}}

\put(55,22.5){\line(1,0){20}}
\put(75,2.5){\line(0,1){20}}

\put(60,20){\line(2,1){10}}
\put(70,10){\line(2,1){10}}

\end{picture}

\caption{The division of (hyper)cube. 
\label{Fig4}}
\end{figure}
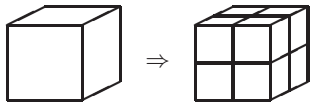

We may extend the model (\ref{DLQG4}) 
in higher space-time dimensions by replacing 
the plaquettes with cubes or hyper-cubes. 
The inhomogeneous lattice could be obtained by replacing one (hyper)cube 
with $2^d$ (hyper)cubes for the model corresponding to the $d$ dimensional 
space-time as in Fig.~\ref{Fig4}. 
Then, as in eq.(\ref{DLQG3}), we may consider the partition function $Z=x+z$ 
by identifying the number of the (hyper)cubes as the volume of the manifold. 
Instead of eq.(\ref{DLQG4}), we obtain,
\be
\label{DLQG15}
z= \left( x + z \right)^{2^d}\, .
\ee
By solving the above algebra equation, we can find the partition function $Z$. 
Eq.~(\ref{DLQG15}) tells that, as in the two dimensional case, we find that there is 
a critical value of $x$ corresponding to Eq.~(\ref{DLQG8}) and 
there is a continuum limit.

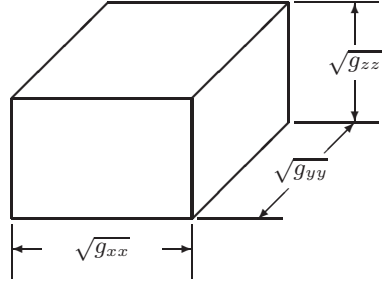
\begin{figure}[h!]
\unitlength=0.8mm
\begin{picture}(90,60)
\thicklines

\put(0,10){\line(1,0){30}}
\put(0,10){\line(0,1){20}}
\put(0,30){\line(1,0){30}}
\put(30,10){\line(0,1){20}}
\put(16,46){\line(1,0){30}}
\put(46,26){\line(0,1){20}}
\put(0,30){\line(1,1){16}}
\put(30,30){\line(1,1){16}}
\put(30,10){\line(1,1){16}}

\thinlines
\put(0,0){\line(0,1){9}}
\put(30,0){\line(0,1){9}}
\put(31,10){\line(1,0){14}}
\put(47,26){\line(1,0){14}}
\put(47,46){\line(1,0){14}}

\put(5,5){\vector(-1,0){5}}
\put(25,5){\vector(1,0){5}}
\put(15,5){\makebox(0,0){$\sqrt{g_{xx}}$}}

\put(46,15){\vector(-1,-1){5}}
\put(52,21){\vector(1,1){5}}
\put(48,18){\makebox(0,0){$\sqrt{g_{yy}}$}}

\put(57,33){\vector(0,-1){7}}
\put(57,39){\vector(0,1){7}}
\put(57,36){\makebox(0,0){$\sqrt{g_{zz}}$}}

\end{picture}

\caption{Cuboid corresponding to the metric in three dimensions. 
}\label{Fig5}
\end{figure}

We should note, however, that the model could not describe the gravity 
although the model itself might be interesting.
In two dimensions, we can choose the conformal gauge as in eq.(\ref{DLQG2}), 
where the conformal factor is the only variable, which allows us to 
identify the number of plaquettes as the area of the manifold. 
In three dimensions, we can choose the gauge condition where the off-diagonal 
components of the metric vanish, 
\be
\label{DLQG16}
ds^2 = g_{xx} dx^2 + g_{yy} dy^2 + g_{zz} dx^2 \, .
\ee
Then we may identify the length of the links, which are the edges of the cuboid 
in the lattice, with the root of the diagonal components of the metric. 
As an example, the square of the length of the link in the $x$ direction could be 
equal to $g_{xx}$ as in Fig.~\ref{Fig5}. 
In four or more than four dimensions, however, we cannot gauge away all 
the off-diagonal components in the metric, and we might need to use a lattice 
whose plaquettes are parallelograms. 
Furthermore in (\ref{DLQG15}), we only take account of the volume of the 
manifold. 
Then we need to define the scalar curvature in order to include 
the Einstein-Hilbert term, which has a dynamical meaning in the dimensions 
higher than two. 

\section{Conclusions}

In summary, we have proposed a new formulation of the discretized quantum 
gravity. In two dimensions, this formulation could surely give a toy model of the 
quantum gravity. 
The continuum limit can be defined in any dimensions and we can calculate 
some physical quantities. 
We have regarded the model in two dimensions as a model of the quantum 
gravity in two dimension but it might be possible to use the model as an 
effective model of the fluctuation of the horizon in four dimensional 
black hole. 

\section*{Acknowledgments.}

This work is supported by 
MEXT KAKENHI Grant-in-Aid for Scientific Research on Innovative Areas ``Cosmic Acceleration''  (No. 15H05890) (S.N.). One of the authors (R.M.) would like to thank G.~Urbina for the recursive formula (\ref{plaquettes}) and fruitful discussions.

\appendix

\section{Counting of $A_n$ \label{appendix}}

We may count $A_n$ in (\ref{DLQG3}) as follows. 
At each step, one of the plaquettes is divided into 4 new plaquettes, 
with a net increment of three plaquettes.
Then at the $s$-th step, $f(s)=A_{3s+1}$ gives the number of different ways 
$s$ steps can be distributed in four plaquettes 
(The extension to the case that the number of plaquettes is the square of an integers 
is straightforward). 
This can be estimated by using the following formula, 
\be
\label{plaquettes}
f(s) = \sum_{i=1}^4 \left( \binom{4}{i} \sum_{{a}_k \in P_i(s-1)} 
\prod_{j=1}^{i} f(a_{kj})\right) \, ,
\ee
where we set $f(1) = 1$.  
In (\ref{plaquettes}), $P_i(s-1)$ is the set of $k$ distinguishable partitions 
$a_k=\left\{a_{k1},a_{k2}, \dots a_{ki} \right\}$ 
(taking into account the ordering for different numbers) 
of the number $s-1$ in $i$ parts, 
that is $s-1 = a_{k1} + a_{k2} + \dots + a_{ki} \quad \forall k$. 
For example,
\be
\label{Pin-1}
P_2(6)= \left\{ \left\{5,1\right\} , \left\{1,5\right\}, \left\{4,2\right\} , 
\left\{2,4\right\} , 
\left\{3,3\right\} \right\} \, . 
\ee
At the $s$-th step, each of diagrams includes $3s+1$ plaquettes. 
Then the formula (\ref{plaquettes}) can be understood from (\ref{DLQGf}) 
by extracting the diagrams including  $3s+1$ plaquettes as in Fig.~\ref{Fig7}. 
By using (\ref{plaquettes}), we obtain 
\begin{align}
\label{Ans}
f(2) &= \binom{4}{1} f(1) = 4\, , \quad f(3) = \binom{4}{1} f(2) 
+ \binom{4}{2} f(1)^2 = 22 \, , \nn
f(4) &= \binom{4}{1} f(3) + \binom{4}{2} \left( f(2) f(1) + f(2) f(1) \right) 
+ \binom{4}{3} f(1)^3  = 140 \, , \nn
f(5) &= 969 \, , \quad  f(6) =  7084 \, , \quad \dots \, .
\end{align}
In fact, by solving Eq.~(\ref{DLQG4}) recursively, we obtain,  
\begin{align}
\label{DLQG4A}
z =& x^4 + 4 x^3 z + 6 x^2 z^2 +4 x z^3 + z^4 = \left( x + z \right)^4 \nn
=& x^4 + 4 x^3\left( x^4 + 4 x^3 z \right) + 6 x^2 \left( x^4 \right)^2 
+ \mathcal{O}\left( x^{13} \right) \nn
=& x^4 + 4 x^3\left( x^4 + 4 x^3 x^4 \right) + 6 x^2 \left( x^4 \right)^2 
+ \mathcal{O}\left( x^{13} \right) \nn
=& x^4 + 4 x^7 + 22 x^{10} + \mathcal{O}\left( x^{13} \right) \, ,
\end{align}
which reproduces (\ref{Ans}) exactly. 
Indeed, a simple closed formula can be obtained for $A_s$ \cite{A_s}
\be
f(s) =  \frac{1}{4s+1} \binom{4s+1}{s}\, ,
\ee
which provides the enumeration of the vertices of rooted, ordered, incomplete 
quartic trees. In the $s \gg 1$ limit one has
\be
f(s) \simeq \frac{1}{3s+1} \left( \frac{2^{s+1}}{3 \pi s} \right)^{\frac{1}{2}}
\left( \frac{4}{3} \right)^{3s}\, .
\ee
\begin{figure}[h!]

\unitlength=0.5mm
\begin{picture}(260,150)
\thicklines

\put(2,122){\line(1,0){16}}
\put(2,122){\line(0,1){16}}
\put(2,138){\line(1,0){16}}
\put(18,122){\line(0,1){16}}

\put(10,130){\makebox(0,0){$4$}}

\put(30,130){\makebox(0,0){$f(4)$}}

\put(42,130){\makebox(0,0){$=$}}

\put(62,112){\line(1,0){16}}
\put(62,112){\line(0,1){16}}
\put(62,128){\line(1,0){16}}
\put(78,112){\line(0,1){16}}
\put(62,120){\line(1,0){16}}
\put(70,112){\line(0,1){16}}

\put(66,116){\makebox(0,0){$3$}}

\put(62,132){\line(1,0){16}}
\put(62,132){\line(0,1){16}}
\put(62,148){\line(1,0){16}}
\put(78,132){\line(0,1){16}}
\put(62,140){\line(1,0){16}}
\put(70,132){\line(0,1){16}}

\put(66,144){\makebox(0,0){$3$}}

\put(82,112){\line(1,0){16}}
\put(82,112){\line(0,1){16}}
\put(82,128){\line(1,0){16}}
\put(98,112){\line(0,1){16}}
\put(82,120){\line(1,0){16}}
\put(90,112){\line(0,1){16}}

\put(94,116){\makebox(0,0){$3$}}

\put(82,132){\line(1,0){16}}
\put(82,132){\line(0,1){16}}
\put(82,148){\line(1,0){16}}
\put(98,132){\line(0,1){16}}
\put(82,140){\line(1,0){16}}
\put(90,132){\line(0,1){16}}

\put(94,144){\makebox(0,0){$3$}}

\put(120,130){\makebox(0,0){$\binom{4}{1} f(3) $}}

\put(52,80){\makebox(0,0){$+$}}

\put(62,62){\line(1,0){16}}
\put(62,62){\line(0,1){16}}
\put(62,78){\line(1,0){16}}
\put(78,62){\line(0,1){16}}
\put(62,70){\line(1,0){16}}
\put(70,62){\line(0,1){16}}

\put(66,66){\makebox(0,0){$1$}}
\put(66,74){\makebox(0,0){$2$}}

\put(62,82){\line(1,0){16}}
\put(62,82){\line(0,1){16}}
\put(62,98){\line(1,0){16}}
\put(78,82){\line(0,1){16}}
\put(62,90){\line(1,0){16}}
\put(70,82){\line(0,1){16}}

\put(66,94){\makebox(0,0){$1$}}
\put(74,94){\makebox(0,0){$2$}}

\put(82,62){\line(1,0){16}}
\put(82,62){\line(0,1){16}}
\put(82,78){\line(1,0){16}}
\put(98,62){\line(0,1){16}}
\put(82,70){\line(1,0){16}}
\put(90,62){\line(0,1){16}}

\put(94,66){\makebox(0,0){$1$}}
\put(86,74){\makebox(0,0){$2$}}

\put(82,82){\line(1,0){16}}
\put(82,82){\line(0,1){16}}
\put(82,98){\line(1,0){16}}
\put(98,82){\line(0,1){16}}
\put(82,90){\line(1,0){16}}
\put(90,82){\line(0,1){16}}

\put(94,94){\makebox(0,0){$1$}}
\put(86,86){\makebox(0,0){$2$}}

\put(102,62){\line(1,0){16}}
\put(102,62){\line(0,1){16}}
\put(102,78){\line(1,0){16}}
\put(118,62){\line(0,1){16}}
\put(102,70){\line(1,0){16}}
\put(110,62){\line(0,1){16}}

\put(106,66){\makebox(0,0){$1$}}
\put(114,66){\makebox(0,0){$2$}}

\put(102,82){\line(1,0){16}}
\put(102,82){\line(0,1){16}}
\put(102,98){\line(1,0){16}}
\put(118,82){\line(0,1){16}}
\put(102,90){\line(1,0){16}}
\put(110,82){\line(0,1){16}}

\put(114,94){\makebox(0,0){$1$}}
\put(114,86){\makebox(0,0){$2$}}

\put(122,62){\line(1,0){16}}
\put(122,62){\line(0,1){16}}
\put(122,78){\line(1,0){16}}
\put(138,62){\line(0,1){16}}
\put(122,70){\line(1,0){16}}
\put(130,62){\line(0,1){16}}

\put(126,66){\makebox(0,0){$2$}}
\put(126,74){\makebox(0,0){$1$}}

\put(122,82){\line(1,0){16}}
\put(122,82){\line(0,1){16}}
\put(122,98){\line(1,0){16}}
\put(138,82){\line(0,1){16}}
\put(122,90){\line(1,0){16}}
\put(130,82){\line(0,1){16}}

\put(126,94){\makebox(0,0){$2$}}
\put(134,94){\makebox(0,0){$1$}}

\put(142,62){\line(1,0){16}}
\put(142,62){\line(0,1){16}}
\put(142,78){\line(1,0){16}}
\put(158,62){\line(0,1){16}}
\put(142,70){\line(1,0){16}}
\put(150,62){\line(0,1){16}}

\put(154,66){\makebox(0,0){$2$}}
\put(146,74){\makebox(0,0){$1$}}

\put(142,82){\line(1,0){16}}
\put(142,82){\line(0,1){16}}
\put(142,98){\line(1,0){16}}
\put(158,82){\line(0,1){16}}
\put(142,90){\line(1,0){16}}
\put(150,82){\line(0,1){16}}

\put(154,94){\makebox(0,0){$2$}}
\put(146,86){\makebox(0,0){$1$}}

\put(162,62){\line(1,0){16}}
\put(162,62){\line(0,1){16}}
\put(162,78){\line(1,0){16}}
\put(178,62){\line(0,1){16}}
\put(162,70){\line(1,0){16}}
\put(170,62){\line(0,1){16}}

\put(166,66){\makebox(0,0){$2$}}
\put(174,66){\makebox(0,0){$1$}}

\put(162,82){\line(1,0){16}}
\put(162,82){\line(0,1){16}}
\put(162,98){\line(1,0){16}}
\put(178,82){\line(0,1){16}}
\put(162,90){\line(1,0){16}}
\put(170,82){\line(0,1){16}}

\put(174,94){\makebox(0,0){$2$}}
\put(174,86){\makebox(0,0){$1$}}

\put(228,80){\makebox(0,0){$\binom{4}{2} \left( f(1)f(2) + f(2)f(1) \right)$}}

\put(52,30){\makebox(0,0){$+$}}

\put(62,12){\line(1,0){16}}
\put(62,12){\line(0,1){16}}
\put(62,28){\line(1,0){16}}
\put(78,12){\line(0,1){16}}
\put(62,20){\line(1,0){16}}
\put(70,12){\line(0,1){16}}

\put(66,16){\makebox(0,0){$1$}}
\put(66,24){\makebox(0,0){$1$}}
\put(74,16){\makebox(0,0){$1$}}

\put(62,32){\line(1,0){16}}
\put(62,32){\line(0,1){16}}
\put(62,48){\line(1,0){16}}
\put(78,32){\line(0,1){16}}
\put(62,40){\line(1,0){16}}
\put(70,32){\line(0,1){16}}

\put(66,44){\makebox(0,0){$1$}}
\put(66,36){\makebox(0,0){$1$}}
\put(74,44){\makebox(0,0){$1$}}

\put(82,12){\line(1,0){16}}
\put(82,12){\line(0,1){16}}
\put(82,28){\line(1,0){16}}
\put(98,12){\line(0,1){16}}
\put(82,20){\line(1,0){16}}
\put(90,12){\line(0,1){16}}

\put(94,16){\makebox(0,0){$1$}}
\put(94,24){\makebox(0,0){$1$}}
\put(86,16){\makebox(0,0){$1$}}

\put(82,32){\line(1,0){16}}
\put(82,32){\line(0,1){16}}
\put(82,48){\line(1,0){16}}
\put(98,32){\line(0,1){16}}
\put(82,40){\line(1,0){16}}
\put(90,32){\line(0,1){16}}

\put(94,44){\makebox(0,0){$1$}}
\put(86,44){\makebox(0,0){$1$}}
\put(94,36){\makebox(0,0){$1$}}

\put(120,30){\makebox(0,0){$\binom{4}{3} f(1)^3 $}}

\end{picture}

\caption{Evaluation of $A_{13} = f(4)$. 
\label{Fig7}}

\end{figure}


\begin{thebibliography}{99}

\bibitem{David:1984tx}
F.~David,
Nucl.\ Phys.\ B {\bf 257} (1985) 45.
doi:10.1016/0550-3213(85)90335-9

\bibitem{David:1985nj}
F.~David,
Nucl.\ Phys.\ B {\bf 257} (1985) 543.
doi:10.1016/0550-3213(85)90363-3

\bibitem{David:1988hj}
F.~David,
Mod.\ Phys.\ Lett.\ A {\bf 3} (1988) 1651.
doi:10.1142/S0217732388001975

\bibitem{Douglas:1989ve}
M.~R.~Douglas and S.~H.~Shenker,
Nucl.\ Phys.\ B {\bf 335} (1990) 635.
doi:10.1016/0550-3213(90)90522-F

\bibitem{Gross:1989vs}
D.~J.~Gross and A.~A.~Migdal,
Phys.\ Rev.\ Lett.\  {\bf 64} (1990) 127.
doi:10.1103/PhysRevLett.64.127

\bibitem{Brezin:1990rb}
E.~Brezin and V.~A.~Kazakov,
Phys.\ Lett.\ B {\bf 236} (1990) 144.
doi:10.1016/0370-2693(90)90818-Q

\bibitem{Marinari:1990fa}
E.~Marinari and G.~Parisi,
Phys.\ Lett.\ B {\bf 247} (1990) 537.
doi:10.1016/0370-2693(90)91897-K


\bibitem{Nojiri:1990ud}
S.~Nojiri,
Phys.\ Lett.\ B {\bf 253} (1991) 63.
doi:10.1016/0370-2693(91)91364-2

\bibitem{Nojiri:1990jv}
S.~Nojiri,
Phys.\ Lett.\ B {\bf 252} (1990) 561.
doi:10.1016/0370-2693(90)90484-N

\bibitem{Nojiri:1992zu}
S.~Nojiri,
Mod.\ Phys.\ Lett.\ A {\bf 7} (1992) 2979
doi:10.1142/S0217732392002354
[hep-th/9206086].

\bibitem{Janssen:1976}
H. K.~Janssen,
Z. Phys. B {\bf{23}}(4) (1976) : 377-380.
doi:10.1007/BF01316547

\bibitem{A_s}
``Number of dissections of a polygon: binomial(4n,n)/(3n+1).'', Sequence A002293 
of the On-line Encyclopedia of Integer Sequences, http://oeis.org/

\end{thebibliography}
\end{document}